\documentclass{aastex631}

\usepackage{amsmath}
\usepackage{graphicx}
\usepackage{natbib}

\begin{document}

\title{A giant glitch from the magnetar SGR J1935+2154 before FRB 200428}

\author{M.Y. Ge}
\affiliation{Key Laboratory of Particle Astrophysics, Institute of High Energy Physics, Chinese Academy of Sciences, Beijing 100049, China}

\author{Yuan-Pei Yang}
\affiliation{South-Western Institute For Astronomy Research, Yunnan University, Kunming 650504, China}
\affiliation{Purple Mountain Observatory, Chinese Academy of Sciences, Nanjing 210023, China}

\author{Fangjun Lu}
\affiliation{Key Laboratory of Particle Astrophysics, Institute of High Energy Physics, Chinese Academy of Sciences, Beijing 100049, China}
\affiliation{Key Laboratory of Stellar and Interstellar Physics and School of Physics and Optoelectronics, Xiangtan University, Xiangtan 411105, China}

\author{Shiqi Zhou}
\affiliation{School of Physics and Astronomy, Sun Yat-Sen University, Zhuhai 519082, China}

\author{Long Ji}
\affiliation{School of Physics and Astronomy, Sun Yat-Sen University, Zhuhai 519082, China}

\author{S.N. Zhang}
\affiliation{Key Laboratory of Particle Astrophysics, Institute of High Energy Physics, Chinese Academy of Sciences, Beijing 100049, China}
\affiliation{University of Chinese Academy of Sciences, Chinese Academy of Sciences, Beijing 100049, China}

\author{Bing Zhang}
\affiliation{Nevada Center for Astrophysics, University of Nevada, Las Vegas, NV 89154, USA}
\affiliation{Department of Physics and Astronomy, University of Nevada, Las Vegas, NV 89154, USA}

\author{Liang Zhang}
\affiliation{Key Laboratory of Particle Astrophysics, Institute of High Energy Physics, Chinese Academy of Sciences, Beijing 100049, China}
\author{Pei Wang}
\affiliation{National Astronomical Observatories, Chinese Academy of Sciences, Beijing 100101, China}
\author{Kejia Lee}
\affiliation{Department of Astronomy, Peking University, Beijing 100871, China}
\affiliation{National Astronomical Observatories, Chinese Academy of Sciences, Beijing 100101, China}
\author{Weiwei Zhu}
\affiliation{National Astronomical Observatories, Chinese Academy of Sciences, Beijing 100101, China}
\author{Jian Li}
\affiliation{CAS Key Laboratory for Research in Galaxies and Cosmology, Department of Astronomy, University of Science and Technology of China, Hefei 230026, China}
\affiliation{School of Astronomy and Space Science, University of Science and Technology of China, Hefei 230026, Chin}
\author{Xian Hou}
\affiliation{Yunnan Observatories, Chinese Academy of Sciences, Kunming 650216, China}
\author{ Qiao-Chu Li}
\affiliation{School of Astronomy and Space Science, Nanjing University, Nanjing 210023, China}
\affiliation{Key laboratory of Modern Astronomy and Astrophysics, Nanjing University, Nanjing 210023, China}

\begin{abstract}
Fast radio bursts (FRBs) are short pulses observed in radio frequencies usually originating from cosmological distances. The discovery of FRB 200428 and its X-ray counterpart from the Galactic magnetar SGR J1935+2154 suggests that at least some FRBs can be generated by magnetars. However, the majority of X-ray bursts from magnetars are not associated with radio emission. The fact that only in rare cases can an FRB be generated raises the question regarding the special triggering mechanism of FRBs. Here we report a giant glitch from SGR J1935+2154, which occurred approximately $3.1\pm2.5$\,day before FRB 200428, with $\Delta\nu=19.8\pm1.4$ {\rm $\mu$Hz} and $\Delta\dot{\nu}=6.3\pm1.1$\,pHz s$^{-1}$. The corresponding spin-down power change rate $\Delta\dot\nu/\dot\nu$ is among the largest in all the detected pulsar glitches. The glitch contains a delayed spin-up process that is only detected in the Crab pulsar and the magnetar 1E 2259+586, a large persistent offset of the spin-down rate, and a recovery component which is about one order of magnitude smaller than the persistent one. The temporal coincidence between the glitch and FRB 200428 suggests a physical connection between the two. The internally triggered giant glitch of the magnetar likely altered the magnetosphere structure dramatically in favour of FRB generation, which subsequently triggered many X-ray bursts and eventually FRB 200428 through additional crustal cracking and Alfv\'en wave excitation and propagation.

\end{abstract}

\keywords{FRB --- stars: neutron --- magnetars: general
--- X-rays: individual (SGR J1935+2154)}

\section{Introduction}
SGR~J1935+2154 was discovered when it entered an outburst phase in 2014, which was followed by four major activity episodes in February 2015, May to July 2016, and November 2019 \citep{2016MNRAS.457.3448I,2016MNRAS.460.2008K,2017ApJ...847...85Y,2020ApJ...902L..43L,2020ApJ...893..156L}. Starting from April 27 2020, multiple short  bursts and a burst forest including hundreds of bursts from  SGR~J1935+2154 were detected by multiple space X-ray and Gamma-ray instruments \citep{2020ApJ...904L..21Y,2021ApJ...916L...7K}. Surprisingly, during the outburst, a double-peaked low-luminosity fast radio burst (FRB) from the direction of SGR~J1935+2154 was observed by CHIME \citep{2020Natur.587...54C} and STARE2 \citep{2020Natur.587...59B} at April 28 UTC 14:34:24, which was subsequently named as FRB 200428. The fluence of FRB 200428 recorded by STARE2 \citep{2020Natur.587...59B} is $\sim$1.5~MJy~ms, making the brightness record of radio bursts from Galactic magnetars. At the same time, its X-ray counterpart, a bright X-ray burst, was detected by orbital high energy instruments such as {\sl Insight}-HXMT, INTEGRAL, Konus-wind and AGILE \citep{2021NatAs...5..378L,2020ApJ...898L..29M,2021NatAs...5..372R,2021NatAs...5..401T}. {\sl Insight}-HXMT discovered the double X-ray peaks corresponding to the double radio peaks \citep{2021NatAs...5..378L}, and both {\sl Insight}-HXMT and INTEGRAL localized the X-ray burst as coming from SGR~J1935+2154 \citep{2021NatAs...5..378L,2020ApJ...898L..29M}. This is the first time that a counterpart of an FRB was detected at other wavelengths, which allowed the identification of the origin of an FRB.

However, the mechanism triggering FRB 200428 is not well understood yet. It is widely suggested that FRBs are generated by the magnetospheric activities of magnetars, either triggered internally \citep{2018ApJ...868...31Y,2020ApJ...897....1L,2020MNRAS.498.1397L,Yang2021,2021MNRAS.507.2208W,LiQC22} or externally \citep{2017ApJ...836L..32Z,2020ApJ...890L..24Z,2021Innov...200152G,2020ApJ...897L..40D}. Recently, \cite{2022arXiv221011518Y} report that SGR J1935+2154 experienced one spin-down glitch, followed by FRB-like bursts and a pulsed radio episode. So the timing properties of SGR J1935+2154 around the epoch of FRB 200428 may provide crucial clues to unveil the physical process that triggered the FRB. This work, we only focus the timing properties around FRB 200428 considering \cite{2022arXiv221011518Y} have published their results.

\section{Observations and data analysis}
\subsection{Observations and data reduction}

In this work, the observations from \textit{NICER}, {\sl NuSTAR}, Chandra and XMM-Newton are utilized to study the spin evolution of SGR J1935+2154 as listed in Appendix Table 1.

\textit{NICER} is a payload onboard the International Space Station devoted to the study of neutron stars through soft X-ray timing \citep{2016SPIE.9905E..1HG}. Its X-ray Timing Instrument (XTI) is an aligned collection of 56 X-ray  concentrator optics (XRC) and silicon drift detector (SDD) pairs, which records individual photons with good spectral resolution and time resolution to $\sim0.1$\,$\mu$s relative to the Universal Time. The cleaned events data in 0.8-4.0 keV are used for profile and timing analyses by using the standard \texttt{nicerl2} command with ``$\rm underonly\_range=0-200$". The arrival time of each event to barycentre is corrected via \texttt{barycorr} with coordinates $\alpha=19^{\rm h}34^{\rm m}55^{\rm s}.68$ and $\delta$=21$^{\circ}$53$^{\prime}$48$^{\prime\prime}$.2 \citep{2016MNRAS.457.3448I}. The solar ephemeris for Solar System Barycentre (SSB) correction is DE405.

SGR J935+2154 was observed by {\sl NuSTAR} around May 02 and 11 (OBSID 80602313002, 80602313004), 2020, with corresponding exposure times of 38 and 31\,ks \citep{2013ApJ...770..103H}. In this work, we analyze data from two telescopes on NuSTAR (usually labeled by their focal plane modules, FPMA and FPMB) using \textsc{HEASoft} (version 6.29). We utilize \texttt{nupipline} with {\sl NuSTAR} CALDB v20180312 to create GTIs and select a circular region of radius $180^{\prime\prime}$ centred on the pulsar position to extract the source spectrum. The arrival time of each photon is corrected to SSB with the same solar ephemeris.

We processed the data collected by PN of {\sl XMM-Newton} from 2014 to May, 2020  \citep{2001A&A...365L...1J} using the Science Analysis System (\texttt{SAS}) (v14.0.01) software. The time intervals contaminated by flaring particle background are discarded. Events in a circular region with a radius of $50^{\prime\prime}$ centred on the pulsar position are selected to ensure that all the source events are included. Only PN data are utilized to perform the timing analysis considering the time resolution. The arrival time of each photon is corrected to SSB with the same solar ephemeris.

Chandra observed SGR J1935+2154 four times in 2014 and 2016, with the ObsIDs of 15874, 15875, 17314, and 18884, respectively. The corresponding time resolutions in these observations are, 0.44\,s, 2.85\,ms, 2.85\,ms, and 2.85\,ms. Therefore, these four observations are used for timing analyses, as listed in Table \ref{table:obs}. The data were reprocessed with the Chandra Interactive Analysis of Observations software (CIAO, version 4.14) using the calibration files available in the Chandra CALDB 4.9.6 data base. The scientific products were extracted following the standard procedures, but adopting extraction regions with different sizes in order to properly subtract the underlying diffuse component. For 15874, we extract the events from a circular region with radius of 1.5$^{\prime\prime}$ for the timing analysis. While for observations 15875, 17314 and 18884 at continuous clocking (CC) mode, events in rectangular boxes of 3$^{\prime\prime}$ x 2$^{\prime\prime}$ sides aligned to the CCD readout direction are used to perform timing analysis. For the timing analysis, we applied the Solar system barycentre correction to the photon arrival times with AXBARY.

\subsection{Timing analysis}

We perform both partially phase-coherent timing (PPCT) analysis and fully phase-coherent timing (FPCT) analysis for SGR J1935+2154 to study its timing behaviors using TEMPO2 \citep{2006MNRAS.369..655H}. The PPCT analysis can mitigate the pronounced effects of timing noise on the long term evolution and show spin evolution clearly \citep{2015ApJ...812...95F}, while the FPCT analysis can get more accurate spin parameters of the pulsar, because the timing noise such as glitches revealed by PPCT analysis can be included in the new timing model. To perform PPCT analysis, we need to have spin frequencies and times of arrival (ToAs) in different epochs, which are obtained with $Z^2_1$ searching, i.e., the frequency making the folded profile deviating the most from a uniform distribution  as represented by the $Z^2_{1}$ value is taken as the spin frequency at the time of each observation, and the phase of the minimum point in each profile is then taken as the TOA of that observation \citep{2012ApJS..199...32G,2020ApJ...904L..21Y}.

The data used for timing analysis span in about 2300\,days from MJD 56853 to 59172. Two obvious abnormalities of SGR J1935+2154 are recognised since the frequencies obtained with $Z^2_1$ searching in two epochs deviate obviously from the extrapolation of the earlier data. In order to show the spin evolution clearly, as discussed above, the PPCT analysis is  performed \citep{2015ApJ...812...95F,2019NatAs...3.1122G}, and the resulted spin parameters in different epochs are given in Table \ref{table:part_para}. The spin parameters obtained from the above mentioned time spans are then fitted with equation (\ref{eq00}) in TEMPO2.
\begin{equation}
\nu(t)=\nu_0+\dot\nu_0{(t-t_{0})}
\label{eq00}
\end{equation}
where $\nu_0$, $\dot\nu_0$ are frequency and frequency derivative at the reference time $t_{\rm 0}$, and $t$ corresponds to
the center of each sub-data set.

Based on the above preliminary timing results of SGR J1935+2154, we can further perform FPCT analysis to get more accurate timing solutions using the data around G1 and G2 respectively, with equation (\ref{eq10}) shown below.
\begin{equation}
\nu(t)=\nu_0+\dot\nu_0{(t-t_{0})}+\Delta{\nu}_{\rm p}+\Delta\dot\nu_{\rm p}{(t-t_{\rm g})}+
\Delta{\nu_{\rm d1}}{\rm e}^{-\frac{(t-t_{\rm g})}{\tau_{\rm d1}}}+\Delta{\nu_{\rm d2}}{\rm e}^{-\frac{(t-t_{\rm g})}{\tau_{\rm d2}}}
\label{eq10}
\end{equation}
where $\nu_0$, $\dot\nu_0$ are frequency and frequency derivative at epoch $t_{\rm 0}$, $\Delta\nu_{\rm p}$ and $\Delta\dot\nu_{\rm p}$ are the persistent offsets of frequency and frequency derivative at the glitch epoch $t_{\rm g}$,  $\Delta\nu_{\rm d1}$, $\Delta\nu_{\rm d2}$, $\tau_{\rm d1}$ and $\tau_{\rm d2}$ are the parameters of the two exponential components. The overall amplitudes of the glitches can be then inferred with $\Delta\nu=\Delta\nu_{\rm p}+\Delta\nu_{\rm d1}+\Delta\nu_{\rm d2}$ and $\Delta\dot\nu=\Delta\dot\nu_{\rm p}-\Delta\nu_{\rm d1}/\tau_{\rm d1}-\Delta\nu_{\rm d2}/\tau_{\rm d2}$. Furthermore, the recover factor $Q=\Delta\nu_{\rm d2}/\Delta\nu$ can be calculated as usually defined \citep{2014ApJ...784...37D,2021MNRAS.508.3251L}. The detailed timing parameters for the two glitches are listed in Table \ref{table:para}.
The errors of spin parameters are obtained from TEMPO2 software.

For SGR J1935+2154, the timing results have not been affected by the timing accuracy of different telescopes. For NICER and NuSTAR, their time resolutions are 0.1\,$\mu$s and 0.1\,ms, respectively \cite{2016SPIE.9905E..1HG,2021ApJ...908..184B}, which much shorted the spin period of SGR J1935+2154. The time resolutions of XMM-Newton and Chandra observations are not as good as NICER and NuSTAR but they do not show differences on the timing results as reported in \cite{2016MNRAS.457.3448I}.

\section{Results}

\subsection{The spin evolution}
\begin{figure*}
\begin{center}
\caption{The spin evolution of SGR J1935+2154. Panel (a): The evolution of spin frequency $\nu$ by subtracting the polynomial component as parameterized in Table \ref{table:para} by PPCT method. Panel (b): The evolution of spin-down rate $\dot\nu$. The values of $\nu$ and $\dot\nu$ are obtained according to PPCT analysis and listed in Table \ref{table:part_para}.  Panel (c): Timing residuals of the model fitting with parameters listed in Table \ref{table:para} by FPCT method.  Panels (d), (e) and (f) illustrate the corresponding zoomed in structures of (a), (b) and (c) around FRB 200428. The solid and dashed red lines represent the fitted spin evolution including and not including the delayed spin-up components. The vertical dot-dashed green line represents the epoch of FRB 200428 while the vertical thin dotted lines represent the epochs of G1 and G2. The dotted blue and magenta lines represent the occurrence time of the two weak radio bursts detected by FAST and Westerbork \citep{2020ATel13699....1Z,2021NatAs...5..414K}, respectively. The confidence level for all the data points are 1-$\sigma$ level in this paper.}
\includegraphics[width=0.45\textwidth,clip]{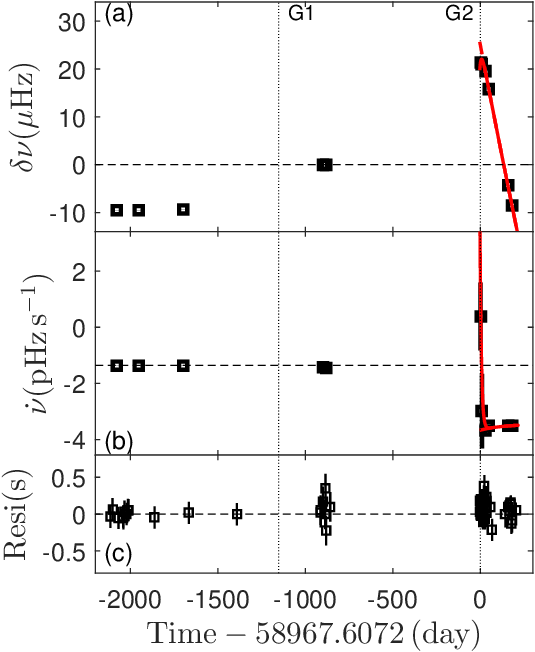}
\includegraphics[width=0.446\textwidth,clip]{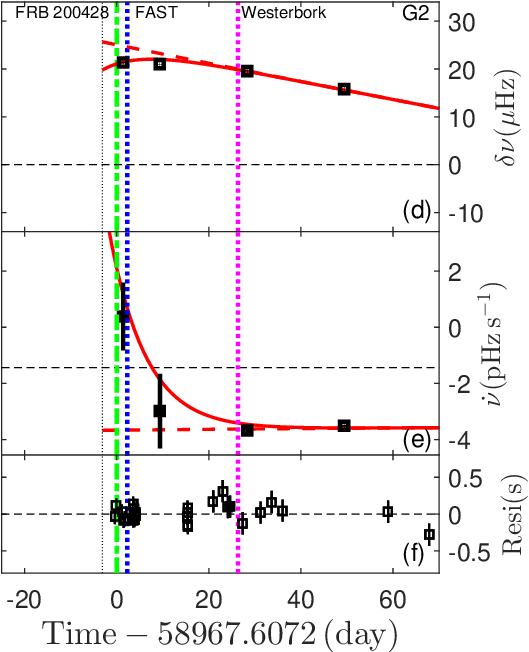}
\label{fig_spin}
\end{center}
\end{figure*}

\begin{figure*}
\begin{center}
\caption{The timing residuals of SGR J1935+2154 with timing parameters pre-G2. Panel (a): the residuals before glitch epoch of G2. Panel (b): residuals with the timing parameters pre G2. The dot dashed line represents the glitch epoch of G2.}
\includegraphics[width=0.6\textwidth,clip]{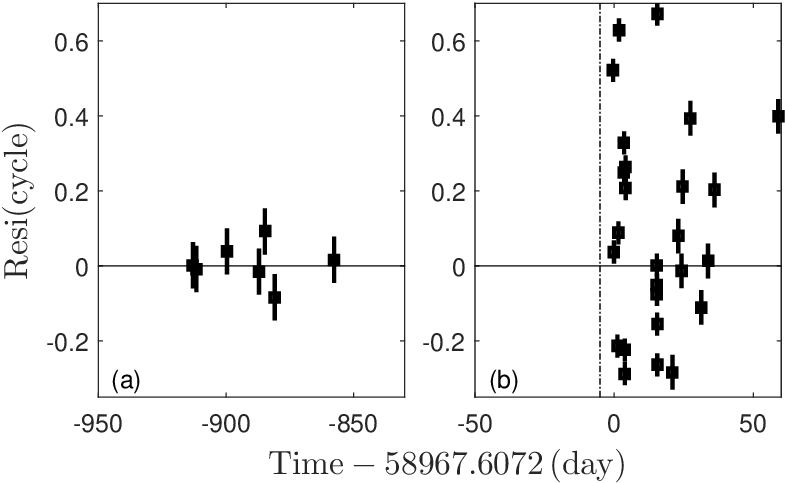}
\label{fig_re}
\end{center}
\end{figure*}

\begin{figure*}
\begin{center}
\caption{The X-ray pulse profiles of SGR J1935+2154 normalised by the mean count rate in different epoches. The red squares and line represent the pulse profile between MJD 58053 and 58110 (2017.10.27-29) obtained with {\sl NICER}. The green squares and line represent the pulse profile obtained with {\sl NICER} between MJD 58967 and 58970 (2020.04.28-05.01), which covers FRB 200428 \citep{2020ApJ...902L...2B}. The blue squares and line represent the pulse profile obtained with {\sl NuSTAR} at MJD 58971 and 58979 (2020.05.02 and 05.10) \citep{2020ApJ...902L...2B,2022arXiv220504983B}. The magenta squares and line are the pulse profile obtained with {\sl XMM-Newton} at MJD 58972 (2020.05.03)  \citep{2022arXiv220504983B}. The cyan squares and line represent the pulse profile between MJD 58986 and 59010 (2020.05.17-06.10) obtained with {\sl NICER}. The pulse profiles are moved upward with a step size of 0.3 to show them more clearly.}
\includegraphics[width=0.4\textwidth,clip]{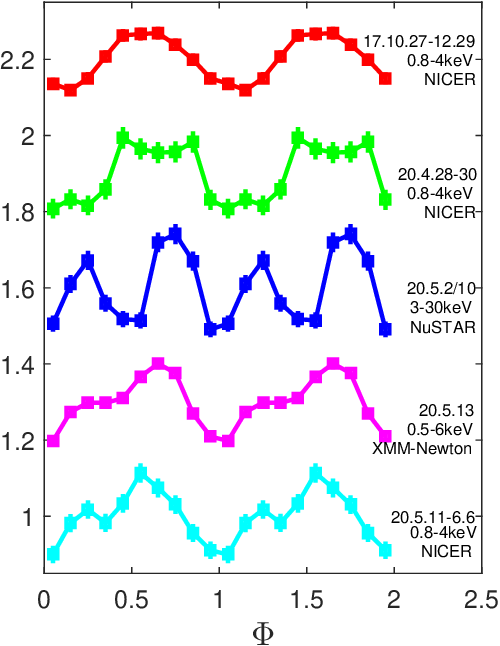}
\label{fig_profile}
\end{center}
\end{figure*}

The long time evolution and short time variations of $\nu$ and $\dot\nu$ could be illustrated directly as in Figure \ref{fig_spin}, which shows two spin-up glitches on MJD 57822 and 58964.5, which are named G1 and G2. The information about the first glitch (G1) is quite limited due to the incomplete time coverage and not concerned in this work. Unfortunately, as no timing information about SGR J1935+2154 is available between MJD 58110 and 58965, we do not know the exact rotation state of SGR J1935+2154 before the burst forest and FRB 200428. However, as shown in Figures \ref{fig_spin} and \ref{fig_re}, the timing behaviors around FRB 200428, which are obtained with both PPCT and FPCT analysis from the observations, are very typical for a glitch, which has three components, the persistent, the delayed spin-up and the recovering components, similar to the big glitches in the Crab pulsar \citep{2020ApJ...896...55G}. The fitting parameters as listed in Table \ref{table:para} show that G2 is a giant glitch happened on MJD 58964.5(2.5) with $\Delta\nu/\nu=(6.4\pm0.4)\times{10}^{-5}$ and $\Delta\dot\nu/\dot\nu=-4.4\pm0.7$, where both $\Delta\nu$ and $\Delta\dot{\nu}$ include the delayed spin-up component that will be discussed later. There are two exponential components following G2. The first exponential component with $\tau_{\rm d1}=8\pm1$\,day represents the delayed spin-up process that was previously only detected in large glitches of the Crab pulsar and magnetar 1E 2259+586 \citep{2020ApJ...896...55G,2004ApJ...605..378W}. The second one is a slow recovery process with $\tau_{\rm d2}=131\pm6$\,day and $\Delta\nu_{\rm d2}=2.50\pm0.16 {\rm {\mu}Hz}$. Remarkably, in the evolution of the spin-down rate, a large persistent offset ($\Delta\dot{\nu}_p$) of $-2.031\pm0.019$\,{\rm pHz s$^{-1}$} is present, which is 1.4 times of the spin-down rate $\dot{\nu}$ before the glitch and is about one order of magnitude larger than the slow recovery component. Such a persistent offset
should be due to an increase in the external torque caused by a rearrangement of the magnetosphere \citep{1992ApJ...390L..21L,2022ApJ...932...11Z}, and the large value implies that the magnetosphere changes dramatically at the glitch, consistent with the large pulse profile changes after G2 \citep{2020ApJ...904L..21Y} (Figure \ref{fig_profile}).

\subsection{Tight constraint on on the epoch of G2}

\begin{figure*}
\begin{center}
\caption{The lightcurve of {\sl NICER} ObsID 3020560101. The light curve of burst forest is removed. The dashed line represents the time of FRB 200428.}
\includegraphics[width=0.6\textwidth,clip]{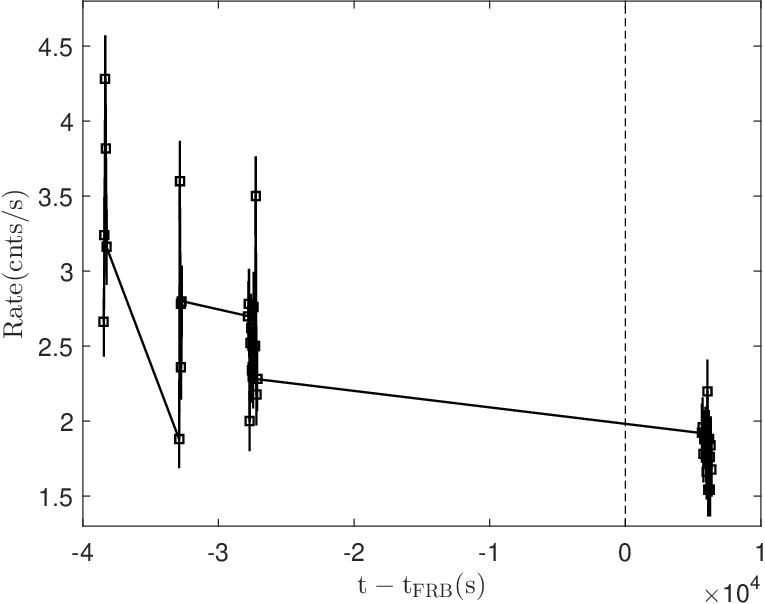}
\label{fig_lc}
\end{center}
\end{figure*}

\begin{figure*}
\begin{center}
\caption{ $Z_{1}^{2}$ variation with rotation frequency of SGR J1935+2154 around FRB 200428 using the {\sl NICER} data. The red, green and blue lines represent the $Z_{1}^{2}$ values versus frequency for the pre-FRB data (35000 to 40000\,s ahead of ObsID 3020560101 as shown in Figure \ref{fig_lc}), the post-FRB data (5500 to 7000\,s of ObsID 3020560101 and ObsIDs 3020560102, 3020560103, 3655010101 and 3655010102) and the whole dataset. The curves show that the frequency remains quite stable in this duration.
}
\includegraphics[width=0.5\textwidth,clip]{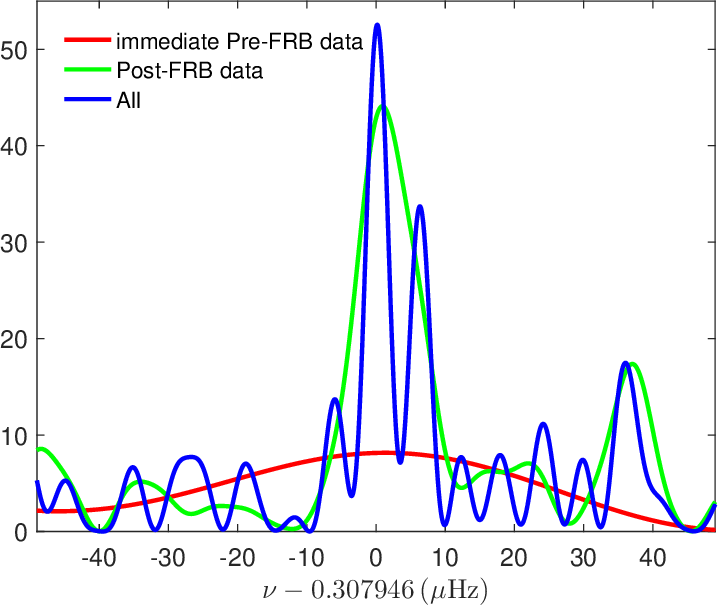}
\label{fig_spin_search}
\end{center}
\end{figure*}

\begin{figure*}
\begin{center}
\caption{The pulse profiles of SGR J1935+2154 obtained with the {\sl NICER} data before and after FRB 200428. The blue points represent the pulse profile before FRB 200428, and the red points represent the pulse profile after FRB 200428. The data selection is the same as in Figure \ref{fig_spin_search}.}
\includegraphics[width=0.5\textwidth,clip]{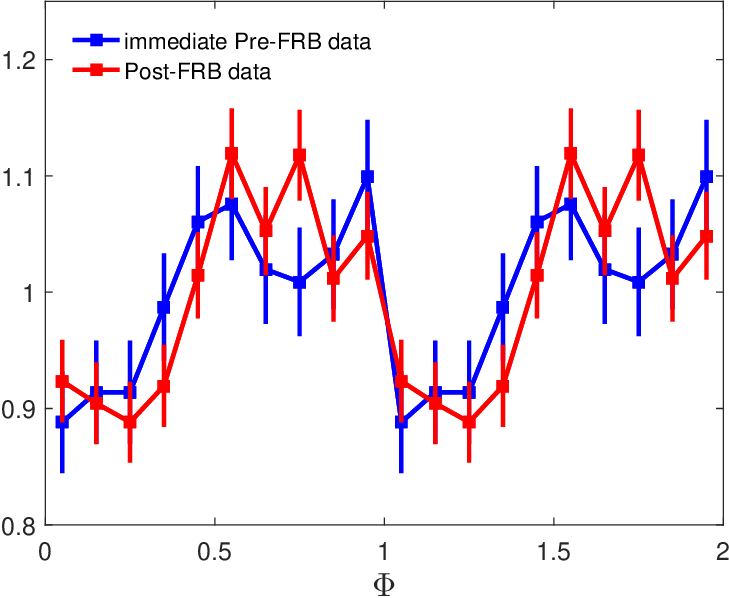}
\label{fig_profile_FRB}
\end{center}
\end{figure*}

The overall timing analyses have not given a very tight constraint on the occurrence time of G2. We therefore discuss whether it happened before FRB 200428 or not with detailed studies on timing behaviors and profiles, since it is crucial for understanding the possible causal connection between the glitch and FRB 200428. The data are divided into two part according to the epoch of FRB 200428 as plotted in Figure \ref{fig_lc}. In Figure \ref{fig_spin_search}, we give the $Z_{1}^{2}$ variations with spin frequency for the pre-FRB data, the post-FRB data and the whole dataset, respectively. The $Z_{1}^{2}$ values for the three datasets all reach their respective maxima at frequency around $0.3079468(12)$\,Hz, and the value of the whole dataset is the highest, which is also consistent with the results from \cite{2020ApJ...904L..21Y} and \cite{2020ApJ...902L...2B}. The difference of peak frequencies is smaller than $1.2\,\mu$Hz, much smaller than $19\,\mu$Hz, the frequency jump of G2, meaning that G2 must happened in advance, i.e., before FRB 200428. We also fold the pre- and post-FRB pulse profiles of this magnetar with the same set of spin parameters obtained above. As shown in Figure \ref{fig_profile_FRB}, the two profiles are not different from each other significantly and share the same minimum phase, supporting the previous conclusion from another aspect. Therefore, G2 happened at least before MJD 58967.2, which is 0.4\,day before FRB 200428. From these two aspects, G2 occurred at least 0.4\,days before FRB 200428 and probably occurred at $3.1\pm2.5$\,day earlier than FRB 200428.

\subsection{Delayed spin-up components}

\begin{figure*}
\begin{center}
\caption{The correlation between $\Delta{\nu}/\nu$ and time scale $\tau_{d1}$ of the delayed spin-up components of glitches in the Crab pulsar, SGR J1935+2154 and 1E 2259+586. The blue squares, red circle and magenta square represent those of the Crab pulsar, SGR J1935+2154 and 1E 2259+586 \citep{2004ApJ...605..378W,2020ApJ...896...55G,2021MNRAS.505L...6S}, respectively.
The green triangle denotes the detected duration upper limit (12.4\,s) of the spin frequency jump for glitches of the Vela pulsar \citep{2018Natur.556..219P}, which is the first for more constraint on the rise time of glitch. The dashed line represents a power-law fit with $\alpha=2.0$.}
\includegraphics[width=0.6\textwidth,clip]{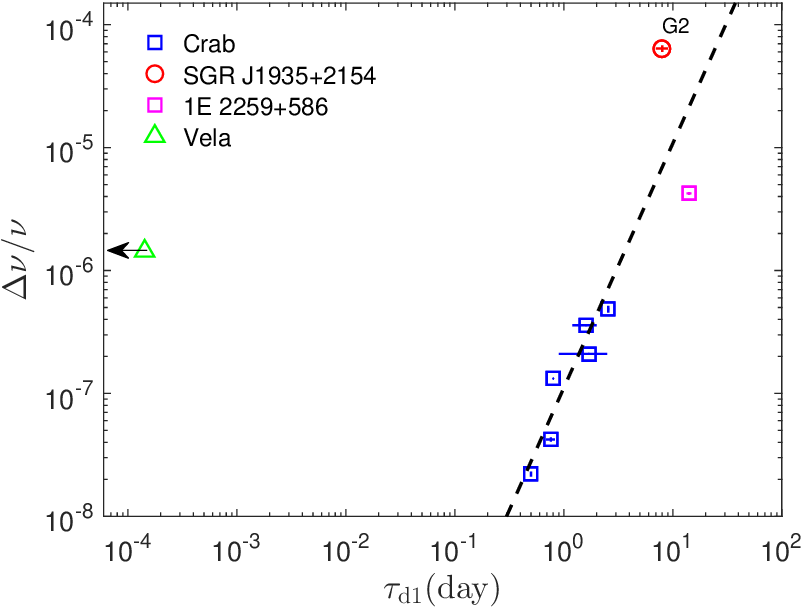}
\label{fig_cor_2psr}
\end{center}
\end{figure*}

We compare the delayed spin-up component of G2 with that of the glitches in the Crab pulsar and 1E 2259+586, as presented in Figure \ref{fig_cor_2psr}. Interestingly, in the $\Delta\nu/\nu$ -$\tau_{\rm d1}$ diagram, where $\tau_{\rm d1}$ is the timescale of the spin-up component, all the spin-up events can be roughly fitted with a power-law function ($\alpha=2.0$). This indicates that the mechanisms for angular momentum transfer of the glitches in the Crab pulsar, 1E 2259+586 and SGR J1935+2154 are similar. The existence of the rather long timescale spin-up components in the Crab pulsar and the non-detection of such components from the Vela pulsar and PSR J0537-6910 are suggested to be due to the different states of their crusts and interiors \citep{2020ApJ...896...55G}, i.e., the Crab pulsar is younger and hence less solidified. The existence of the delayed spin-up component of SGR J1935+2154 thus implies that it is young. It is worth noting here that the delayed spin-up components may have been detected from several other magnetars even though their timing properties are not well resolved \citep{2004ApJ...605..378W,2014ApJ...784...37D}, indicating that they are also young as widely believed.

\subsection{Comparison with other glitch samples}
\begin{figure*}
\begin{center}
\caption{Panel (a): Comparison of glitch G2 in SGR J1935+2154 with other glitches of rotation powered pulsars in the $\Delta{\nu}/\nu-\Delta{\dot\nu}/\dot\nu$ diagram. Panel (b): The $\Delta{\nu}-\Delta{\dot\nu}$ diagram. The red circles, green circle, blue squares and grey dots represent the two glitches of SGR J1935+2154, the anti-glitch of SGR 1900+14 \citep{1999ApJ...524L..55W}, glitches of other known magnetars, and glitches of normal rotation-powered pulsars \citep{2014ApJ...784...37D,2011MNRAS.414.1679E,2021MNRAS.508.3251L}, respectively. The empty blue squares represent the anti-glitches of 1E 2259+586 \citep{2013Natur.497..591A}. The black circles, squares, triangles and stars in panel (b) represent the glitches of the Crab pulsar, the Vela pulsar, PSR J0537-6910 and PSR J0205+6449 \citep{2014ApJ...784...37D,2021MNRAS.508.3251L}, respectively. The glitch catalog also be obtained from website: http://www.jb.man.ac.uk/pulsar/glitches/gTable.html and
https://www.atnf.csiro.au/people/pulsar/psrcat/glitchTbl.html.
}
\includegraphics[width=0.4\textwidth,clip]{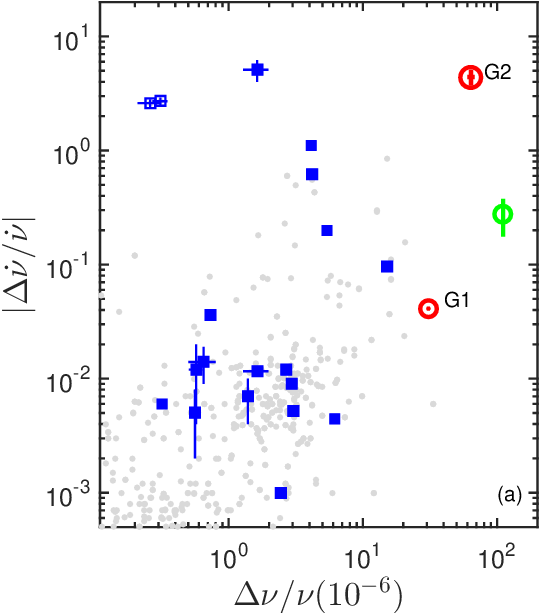}
\includegraphics[width=0.4\textwidth,clip]{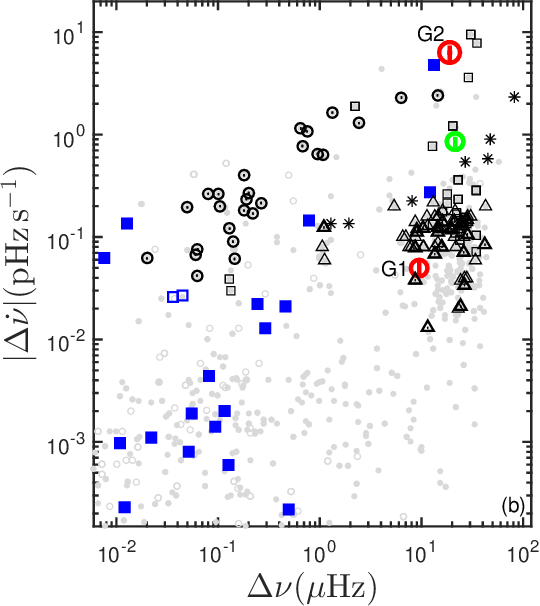}
\label{fig_f0f1_dis}
\end{center}
\end{figure*}

From the known glitch sample of isolated pulsars, this event of SGR J1935+2154 is among the largest two in terms of the relative glitch amplitudes in the $\Delta\nu/\nu$-$\Delta\dot\nu/\dot\nu$ diagram \citep{2014ApJ...784...37D,2011MNRAS.414.1679E,2021MNRAS.508.3251L} as shown in Figure \ref{fig_f0f1_dis}. The other glitch-like event in magnetars with a comparable relative amplitude is associated with the super burst of SGR 1900+14 around MJD 51050 \citep{1999ApJ...524L..55W}. Although the $\Delta\nu$ and $\Delta\dot\nu$ values of the large glitches in the Crab pulsar, Vela pulsar, PSR J0205+6449 and PSR J0537-6910 are comparable with those of G2 in SGR J1935+2154 (as shown in Figure \ref{fig_f0f1_dis} (b)), the relative change of the rotation frequency is much greater than those of the pulsars considering that this magnetar rotates much more slowly than the pulsars. Moreover, among the magnetar glitch sample, G2 has the largest $\Delta\nu$ value while SGR 1900+14 has the largest  $\Delta\nu<0$ value \citep{1999ApJ...524L..55W}.

\section{Discussion}

For the physical connection between FRBs and glitches, we suggest the following scenario based on the previous theoretical framework and the observational properties of SGR J1935+2154 around the glitch. After the glitch, the motion of the core superfluid neutron vortices in the direction perpendicular to the spin axis during the spin-down relaxation phase of the glitch would alter the core magnetic field, which would result in the movement of the neutron star crust and the change in the surface magnetic field \citep{Ruderman1998}. Crustquakes are expected if the solid crust does not plastically adjust to the less-oblate equilibrium shape required by the pulsar's spin-down or if the magnetic stress exceeds the shear modulus \citep{Baym1971,Thompson1995,Perna2011,LiQC22}. Crustal fracturing produces Alfv\'en waves and then generates X-ray bursts \citep{Thompson1995}.
On the other hand, some FRB emission models proposed that FRB are associated with magnetar activity, such as FRBs produced in the charge starvation region triggered by Alfv\'en waves or crust plate motions \citep{Kumar2020,2020MNRAS.498.1397L,Wadiasingh20,Yang2021}, FRB produced by relativistic plasmoids/outflows lunched by Alfv\'en waves \citep{Metzger19,Yuan20}, FRB generated by magnetic reconnection near the light-cylinder \citep{Lyubarsky20}, etc.
Immediately after the glitch, frequent magnetic activities generate plenty of energetic charged particles in the form of electron-positron pairs, which easily power X-ray bursts. However, if FRBs are produced in the magnetosphere as proposed by some models \citep[e.g.,][]{Kumar2020,2020MNRAS.498.1397L,Wadiasingh20,Yang2021}, they would be difficult to be generated in such a pair-rich environment due to the following two reasons: 1) the abundant pairs would shield the charge starvation region necessary for FRB generation; 2) even if an FRB is generated under some specific conditions, it is difficult to escape from a pair-rich magnetosphere due to a large scattering optical depth in the magnetosphere. Some time later, the magnetic field configuration becomes less irregular, as represented by the decrease of X-ray burst frequency \citep{2022ApJS..260...24C}. Particles can escape easily from the magnetosphere in the form of a pulsar wind like that in PSR B0540--69 after a spin-down rate transition \citep{2019NatAs...3.1122G}. FRBs and weak radio bursts may be then generated due to the formation of the charge starvation region in the magnetosphere. This explains why FRB 200428 occurred a few days after G2 \citep{2020ATel13699....1Z,2021NatAs...5..414K} . Long after the glitch, as the crust plate is recovered and magnetic field rearrangement is finished, magnetic activities become less frequent and FRBs become more difficult to be generated. In short, FRBs preferably occur some time after the most active episodes triggered by giant glitches in magnetars. It is therefore essential to closely monitor the spin evolution of magnetars in X-rays and coordinate radio follow-up observations when a major glitch is detected in order to further test this FRB generation scenario.

\section{Summary}

We utilize the observations from \textit{NICER}, {\sl NuSTAR}, Chandra and XMM-Newton to study the timing behaviour of SGR J1835+2154 and the possible trigger mechanism of FRB 200428. From the timing result, we find a giant glitch from SGR J1935+2154, which occurred approximately $3.1\pm2.5$\,day before FRB 200428, with $\Delta\nu=19.8\pm1.4$ {\rm $\mu$Hz} and $\Delta\dot{\nu}=6.3\pm1.1$\,pHz s$^{-1}$. The corresponding spin-down power change rate $\Delta\dot\nu/\dot\nu$ is among the largest in all the detected pulsar glitches. The glitch contains a delayed spin-up process that is only detected in the Crab pulsar and the magnetar 1E 2259+586, a large persistent offset of the spin-down rate, and a recovery component which is about one order of magnitude smaller than the persistent one. The temporal coincidence between the glitch and FRB 200428 suggests a physical connection between the two, which supplies more constraint on the trigger mechanism of FRB 200428.

\clearpage


\clearpage

\clearpage



\begin{table}
\small
\caption{The spin parameters of SGR J1935+2154 obtained with partially phase-coherent timing analysis.}
\label{table:part_para}
\medskip
\begin{center}
\begin{tabular}{l l l l l l }
\hline \hline
No. &  Start & Finish & PEPOCH & $\nu$ & $\dot\nu$  \\
    &   MJD  &  MJD   &  MJD   &  Hz & ${\rm pHzs^{-1}}$\\
\hline
1 &  56822	   &  56950      &  56899   &  0.308163446(2)   & -1.359(2)  \\
2 &	 56892     &  57335      &  57101   &  0.3081397240(3)  & -1.35892(7)  \\
3 &	 56926     &  57605      &  57251   &  0.3081221081(3)  & -1.35923(5)  \\
\hline
4 &	 58046     &  58086      &  58068   &  0.30803445(2)    & -1.42(9)  \\
5 &	 58067     &  58130      &  58088   &  0.30803204(3)    & -1.44(5)  \\
\hline
6 &	 58966     &  58973      &  58969   &  0.30794566(9)    & 0.38(1.20) \\
7 &	 58970     &  58984      &  58977   &  0.30794430(3)    & -2.99(1.32)\\
8 &	 58987     &  59007      &  58996   &  0.30794063(3)    & -3.67(18)  \\
9 &	 58998     &  59041      &  59017   &  0.307934203(8)   & -3.50(3)   \\

\hline
\end{tabular}
\end{center}
\end{table}

\begin{table}
\small
\caption{The timing parameters of the two Glitches in SGR J1935+2154.}
\label{table:para}
\medskip
\begin{center}
\begin{tabular}{llll}
\hline \hline
Parameters                               & G1 & G2$^{*}$      \\
\hline
Epoch (MJD)                              & 57214 & 58088   \\
$\nu $(${\rm Hz}$)                       & 0.3081264533(3)& 0.30803203(2) \\
$\dot\nu$(${\rm pHzs^{-1}}$)             & -1.35922(4)    &  -1.415(19) \\
\hline
Glitch epoch (MJD)                       & 57822(22)      &  58964.5(2.5)** \\
$\Delta\nu({\rm {\mu}Hz})$               & 9.5(2)         &  19.8(1.4)  \\
$\Delta\nu_{p}({\rm {\mu}Hz})$           & --             &  23.2(1.4)  \\
$\Delta\dot\nu({\rm pHzs^{-1}})$         &-0.0558(0.0123) &  6.3(1.1)   \\
$\Delta\dot\nu_{\rm p}({\rm pHzs^{-1}})$ & --             &  -2.031(19) \\
$\Delta\nu_{\rm d1}({\rm {\mu}Hz})$      & --             &  -5.88(9)   \\
$\tau_{\rm d1}({\rm d})$                 & --             &  8(1)       \\
$\Delta\nu_{\rm d2}({\rm {\mu}Hz})$      & --             &  2.50(16)   \\
$\tau_{\rm d2}({\rm d})$                 & --             &  131(6)     \\
$\Delta\nu/\nu(10^{-6})$                 & 30.8(9)        &  64(4)      \\
$\Delta\dot\nu/\dot\nu$                  & 0.041(9)       &  -4.4(7)    \\
$Q$                                      & --             &  0.13(1)    \\
Time range                               & 56822--58130   & 58046--59120\\
Residuals (ms)                           & 90.9           &  105.6    \\
\hline
\end{tabular}
\end{center}
*: In the analyses we assume that the timing behaviors between MJD 58110 and 58965 follow the trend in MJD 58054-58110.

**: The uncertainty of the occurrence time of G2 is 2.5\,day. But as discussed in the text, it happened definitely before MJD 58967.2, which is 0.4\,day before FRB 200428 at MJD 58967.60857593.
\end{table}

\clearpage

\section*{Acknowledgments}
This work is supported by the National Key R\&D Program of China (2021YFA0718500) from the Minister of Science and Technology of China (MOST). The authors thank supports from the National Natural Science Foundation of China under Grants U1938109, U1838201, U1838202, 12173103, 12003028, U2038101, U1938103 and 11733009. This work is also supported by International Partnership Program of Chinese Academy of Sciences (Grant No.113111KYSB20190020), SKA Fast Radio Burst and High-Energy Transients Project (2022SKA0130101), and the China Manned Spaced Project (CMS-CSST-2021-B11).

\bibliographystyle{aasjournal}
\bibliography{ms}{}

\clearpage

\section*{Appendix}

\begin{center}
\begin{longtable}{l l l l l}
\setcounter{table}{1}
\\
\caption{The observation catalogue of SGR J1935+2154.
The time resolutions of Chandra observations 15874, 15875, 17314, and 18884 are 0.44\,s, 2.85\,ms, 2.85\,ms, and 2.85\,ms, and those of XMM-Newton and NICER are 68.7\,ms and 1\,$\mu{\rm s}$, respectively.}
\label{table:obs}
\\\hline \hline
Telescope       &  ObsID & MJD & Exposure  \\
                &       &       & ks        \\
\hline
Chandra       &  15874 & 56853  & 10  \\
Chandra       &  15875 & 56866  & 75   \\
Chandra       &  17314 & 56900  & 29   \\
Chandra       &  18884 & 57576  & 20   \\
\hline
XMM-Newton    & 0722412501 & 56926 & 22  \\
XMM-Newton    & 0722412601 & 56928 & 23  \\
XMM-Newton    & 0722412701 & 56934 & 22  \\
XMM-Newton    & 0722412801 & 56945 & 23  \\
XMM-Newton    & 0722412901 & 56954 & 10  \\
XMM-Newton    & 0722413001 & 56957 & 17  \\
XMM-Newton    & 0748390801 & 56976 & 22  \\
XMM-Newton    & 0764820101 & 57106 & 46  \\
XMM-Newton    & 0764820201 & 57302 & 68  \\
XMM-Newton    & 0871190201 & 58982 & 51  \\
\hline
NuSTAR       & 80602313002	& 58971 & 37  \\
NuSTAR       & 80602313004	& 58979 & 38  \\
\hline
NICER	&	1020560101	&	58053		&	3.0	  \\
NICER	&	1020560102	&	58054		&	13.9  \\
NICER	&	1020560103	&	58055		&	8.1	  \\
NICER	&	1020560104	&	58056		&	5.8	  \\
NICER	&	1020560105	&	58057		&	0.4	  \\
NICER	&	1020560106	&	58058		&	2.3	  \\
NICER	&	1020560107	&	58079		&	1.9	  \\
NICER	&	1020560108	&	58080		&	6.4	  \\
NICER	&	1020560109	&	58081		&	2.7	  \\
NICER	&	1020560110	&	58082		&	5.3	  \\
NICER	&	1020560111	&	58084		&	2.9	  \\
NICER	&	1020560112	&	58087		&	1.7	  \\
NICER	&	1020560113	&	58088		&	0.8	  \\
NICER	&	1020560115	&	58115		&	0.9	  \\
NICER	&	1020560116	&	58116		&	0.7	  \\
NICER	&	2020560101	&	58763		&	0.3	  \\
NICER	&	2020560102	&	58764		&	2.2	  \\
NICER	&	2020560103	&	58765		&	2.3	  \\
NICER	&	2020560104	&	58794		&	1.7	  \\
NICER	&	3020560101	&	58967		&	3.1	  \\
NICER	&	3020560102	&	58968		&	0.9	  \\
NICER	&	3655010101	&	58968		&	0.8	  \\
NICER	&	3655010102	&	58969		&	3.9	  \\
NICER	&	3020560103	&	58969		&	0.7	  \\
NICER	&	3020560104	&	58980		&	0.9	  \\
NICER	&	3655010201	&	58987		&	4.7	  \\
NICER	&	3020560105	&	58988		&	0.9	  \\
NICER	&	3020560106	&	58989		&	0.6	  \\
NICER	&	3020560107	&	58991		&	5.2	  \\
NICER	&	3020560108	&	58992		&	3.2	  \\
NICER	&	3020560109	&	58994		&	0.8	  \\
NICER	&	3020560110	&	58997		&	1.7	  \\
NICER	&	3020560111	&	58998		&	1.0	  \\
NICER	&	3020560112	&	58999		&	1.3	  \\
NICER	&	3020560113	&	59000		&	1.1	  \\
NICER	&	3020560114	&	59001		&	2.7	  \\
NICER	&	3020560115	&	59002		&	1.0	  \\
NICER	&	3020560116	&	59003		&	0.7	  \\
NICER	&	3020560117	&   59004		&	1.4	  \\
NICER	&	3020560118	&	59005		&	1.2	  \\
NICER	&	3020560119	&	59006		&	1.0	  \\
NICER	&	3655010301	&	59017		&	1.2	  \\
NICER	&	3655010302	&	59018		&	6.6	  \\
NICER	&	3655010303	&	59018		&	4.1	  \\
NICER	&	3020560120	&	59020		&	0.9	  \\
NICER	&	3020560121	&	59021		&	0.8	  \\
NICER	&	3020560122	&	59022		&	1.7	  \\
NICER	&	3020560123	&	59023		&	3.2	  \\
NICER	&	3020560124	&	59024		&	1.8	  \\
NICER	&	3020560125	&	59025		&	1.7	  \\
NICER	&	3020560126	&	59027		&	0.8	  \\
NICER	&	3020560127	&	59028		&	1.7	  \\
NICER	&	3020560128	&	59029		&	1.7	  \\
NICER	&	3020560129	&	59030		&	2.2	  \\
NICER	&	3020560130	&	59031		&	0.6	  \\
NICER	&	3020560131	&	59032		&	1.2	  \\
NICER	&	3020560132	&	59033		&	1.1	  \\
NICER	&	3020560133	&	59034		&	0.8	  \\
NICER	&	3020560134	&	59038		&	0.5	  \\
NICER	&	3020560135	&	59040		&	0.4	  \\
NICER	&	3020560136	&	59041		&	1.3	  \\
NICER	&	3020560137	&	59042		&	0.7	  \\
NICER	&	3020560138	&	59045		&	3.1	  \\
NICER	&	3020560139	&	59046		&	1.4	  \\
NICER	&	3020560140	&	59047		&	0.9	  \\
NICER	&	3020560141	&	59049		&	0.8	  \\
NICER	&	3020560142	&	59056		&	0.2	  \\
NICER	&	3020560143	&	59061		&	1.3	  \\
NICER	&	3020560144	&	59062		&	1.4	  \\
NICER	&	3020560145	&	59063		&	0.5	  \\
NICER	&	3020560146	&	59064		&	2.0	  \\
NICER	&	3020560147	&	59065		&	0.8	  \\
NICER	&	3020560148	&	59067		&	2.2	  \\
NICER	&	3020560149	&	59089		&	0.3	  \\
NICER	&	3020560150	&	59102		&	0.1	  \\
NICER	&	3020560151	&	59103		&	2.1	  \\
NICER	&	3020560152	&	59111		&	1.2	  \\
NICER	&	3020560153	&	59117		&	1.5	  \\
\hline
\end{longtable}

\end{center}

\end{document}